\documentclass[a4paper,12pt,notitlepage]{article}
\usepackage{latexsym,amsthm,amsfonts,amssymb,amsmath,amsthm}
\usepackage{fullpage}
\usepackage{url}
\usepackage{amsmath,amssymb,amsthm}
\usepackage{setspace,newclude}
\usepackage{hyperref}
\usepackage{graphicx,enumerate}
\usepackage{empheq}
\usepackage[latin1]{inputenc}
\usepackage{tikz}
\usetikzlibrary{shapes,arrows}
\usepackage{fancyhdr}
\usepackage[english]{babel}
\usepackage{cite}
\usepackage{titling}
\newtheorem{theorem}{Theorem}
\newtheorem{lemma}{Lemma}
\newtheorem{definition}{Definition}

\newtheorem{fact}{Fact}

\def\NN{{\bf N}}

\def\FF{{\bf F}}

\def\code{{\mathcal C}}

\def\random{{\mathcal R}}
\def\VV{{\mathcal V}}
\def\PP{{\mathcal P}}

\def\Tree{{\mathcal T}}

\def\polylog{\operatorname{polylog}}

\newcommand{\Ix}{\mbox{\sc Index}}
\newcommand{\OR}{\mbox{\sc OR}}
\newcommand{\NC}{\cal NC}

\pdfoutput=1
\begin{document}
\title{An Improved Interactive Streaming Algorithm for the Distinct Elements Problem}
\author{Hartmut Klauck\thanks{This work is funded by the Singapore Ministry of Education (partly through the Tier 3 Grant "Random numbers from quantum processes") and by the Singapore National Research Foundation.}
\\
CQT and Nanyang Technological University\\
{\tt hklauck@gmail.com}
\and
Ved Prakash\\ 
CQT and National University of Singapore \\
{\tt prakash18ved@gmail.com}
}
\date{}
\maketitle
\begin{abstract}
The exact computation of the number of distinct elements (frequency moment $F_0$) is a fundamental problem in the study of data streaming algorithms. We denote the length of the stream by $n$ where each symbol is drawn from a universe of size $m$. While it is well known that the moments $F_0,F_1,F_2$ can be approximated by efficient streaming algorithms \cite{ams:frequency}, it is easy to see that exact computation of $F_0,F_2$ requires space $\Omega(m)$.
In previous work, Cormode et al.~\cite{Cormode:2011:verifying_computation} therefore considered a model where the data stream is also processed by a powerful helper, who provides an interactive proof of the result. They gave such protocols with a polylogarithmic number of rounds of communication between helper and verifier for all functions in NC. This number of rounds ($O(\log^2 m)$ in the case of $F_0$) can quickly make such protocols impractical.

 Cormode et al.~also gave a protocol with $\log m +1$ rounds for the exact computation of $F_0$ where the space complexity is $O\left(\log m \log n+\log^2 m\right)$ but the total communication $O\left(\sqrt{n}\log m\left(\log n+ \log m \right)\right)$. They managed to give $\log m$ round protocols with $\polylog(m,n)$ complexity for many other interesting problems including $F_2$, Inner product, and Range-sum, but computing $F_0$ exactly with polylogarithmic space and communication and $O(\log m)$ rounds remained open.

\par In this work, we give a streaming interactive protocol with $\log m$ rounds for exact computation of $F_0$ using $O\left(\log m \left(\,\log n + \log m \log\log m\,\right)\right)$ bits of space and the communication is $O\left( \log m \left(\,\log n +\log^3 m (\log\log m)^2 \,\right)\right)$. The update time of the verifier per symbol received is $O(\log^2 m)$.
%
\end{abstract}
\section{Introduction}
In a seminal work \cite{ams:frequency} Alon, Matias and Szegedy studied the space complexity of both approximating the frequency moments of a data stream and computing them exactly. Streaming algorithms are usually designed to handle large data sets, and the algorithm should be able to process each data element with small time overhead, and should have small working space as well.
 For instance one of the striking results of Alon et al.~is that the second frequency moment $F_2$ can be approximated up to a constant factor using only $O(\log n+\log m)$ space by a randomized algorithm, where $m$ is the size of the universe and $n$ is the stream length. The interested reader is referred to the survey by Muthukrishnan \cite{muthukrishnan_book}. Other problems studied in the data stream model include graph problems like matching and triangle counting \cite{Yossef:reductions_streaming_traingle,Feigenbaum:semistreaming}.
\par It is known that the frequency moments $F_j$ for integer $j>2$ are hard to even approximate by any streaming algorithm, i.e., any streaming algorithm giving a good approximation must have large space \cite{yoseff:datastream}. Motivated by this and the paradigm of cloud computing
one can study new models where a helper/prover is introduced. The hope is that while some problems require a lot of space to solve by an unassisted streaming algorithm, a helper who is not space restricted and sees the stream in the same way as the verifier might not only be able to {\em compute} the result, but be able to {\em convince} the client/verifier of the correctness of that result by providing a proof that can be verified by the client using small space only. In the past few years, there have been numerous papers~\cite{Cormode:2010:streaming_graph_computations,Cormode:2011:verifying_computation,Cormode:2012:
practical_streaming,Chakrabarti:annotations,Chakrabarti_OIP_model,ved_hartmut_itcs13} investigating this idea.

  Thus we have the following scenario: both the prover and client observe the stream. The client, who is severely space restricted, computes a sketch of the data. The prover, having no space restriction, can store the entire stream, compute the answer, and send it to the client. But the prover may not be honest, e.g.~the prover may have incentives to not provide the correct answer. The client uses the sketch of the data to reject wrong claims with high probability. The prover can be thought of as an internet company which offers cloud computing services and operates huge data warehouses. The only formal restriction on the prover is that he cannot predict the future parts of the stream. From now on, we refer to the client as the verifier.
\par Besides many upper bounds provided in the papers cited above, one can also show lower bounds against this model of prover assisted data streaming algorithms. Data streaming protocols  can be simulated by Arthur-Merlin communication protocols, where Merlin is the prover and the data stream input is split across some players, who together constitute the verifier Arthur. Arthur-Merlin communication complexity was first introduced by Babai, Frankl and Simon~\cite{bfs:comm_classes} and was studied in greater detail by Klauck~\cite{Klauck_rect_size_bounds,klack_AM_games_CC}. These lower bounds have been used by Chakrabarti et al.~\cite{Chakrabarti:annotations} to give non-trivial lower bounds on approximating and computing exactly the k-th frequency moments for large enough $k$, in the setting where the proof provided is noninteractive, i.e., the prover provides an ``annotation'' to the data stream that is then verified without further interaction.
Unfortunately analyzing model of interactive proofs with many rounds between prover and verifier in communication complexity seems to be out of reach for current techniques in communication complexity.

\par One of the fundamental problems in data streaming is to compute the number of distinct elements in a data stream, which is the zeroth frequency moment and is denoted by $F_0$. This problem has many application in areas such as query optimization, IP routing and data mining (See the references cited in \cite{Kane_approx_F0} for details). By a simple reduction from the disjointness function \cite{razborov:disj}, it is easy to get a lower bound of $\Omega(m)$ (assuming $m=\theta(n)$) on the streaming complexity of computing the exact number of distinct elements by a data streaming algorithm without a prover. If we require exact $F_0$ and the verifier's space to be sublinear, we have to look at the prover-verifier model.
\par By appealing to Klauck's \cite{Klauck_rect_size_bounds} result on the MA complexity of disjointness, there is a lower bound on $hv=\Omega(m)$ to compute $F_0$ exactly in the online MA model as defined in \cite{Chakrabarti:annotations}, where $h$ is the help cost and $v$ is the space used by the streaming algorithm. Cormode et al.~\cite{Cormode:2011:verifying_computation} gave interactive streaming protocols with $\log m$ rounds for various interesting problems like frequency moments, the Index function and computing Inner Products. They also gave a general purpose protocol that computes every function in NC with polylogarithmic space, communication and rounds.
For the case of exactly computing $F_0$, the general purpose protocol uses $O(\log^2 m)$ rounds, which can very quickly become impractical. Hence the authors also describe a protocol using only $\log m$ rounds, where the help cost (i.e., communication)  is not polylogarithmic in $m$ and $n$. We improve their protocol so that both the communication $h$ and the space $v$ are polylogarithmic in $m$ and $n$, while using only $\log m$ rounds of interaction.
\subsection{Previous Work}
Let $m$ be the universe size and $n$ be the length of the stream. Although we later state our complexity results in terms of $m$ and $n$, \emph{in this subsection, for simplicity, we assume $m$ and $n$ are roughly of the same order of magnitude}, i.e. $m=poly(n)$ following the previous works in \cite{Cormode:2011:verifying_computation,Cormode:2012:practical_streaming}. It is known that approximating $F_0$ up to a $(1\pm\epsilon)$ multiplicative factor can be done in $O(\epsilon^{-2}+\log m)$ space using randomization, which is optimal as well \cite{Kane_approx_F0}.

\par Goldwasser, Kalai and Rothblum \cite{Goldwasser:2008:muggles_paper} proposed a delegation general purpose interactive protocol for log-space uniform $\NC$ circuits. Their protocol was presented formally in the streaming setting by Cormode, Mitzenmacher, and Thaler~\cite{Cormode:2012:practical_streaming}. We state their results below for easy reference.

\begin{fact}\label{fact:muggles} [Theorem~3.1 from \cite{Cormode:2012:practical_streaming}] \\
Let $f$ be a function over an arbitrary field $\FF$ that can be computed by a family of $O(\log S(n))$-space uniform arithmetic circuits(over $\FF$) of fan-in $2$, size $S(n)$ and depth $d(n)$. Then in the streaming model with a prover, there is a protocol which requires $O(d(n) \log S(n))$ rounds such that the verifier needs $O\left(\log S(n) \log |\FF|\right)$ bits of space and the total communication between the prover and the verifier is $O\left(d(n) \log S(n) \log |\FF|\right)$.
\end{fact}
As a result of Fact~\ref{fact:muggles}, if we use the general purpose interactive protocol of \cite{Goldwasser:2008:muggles_paper} to compute $F_0$ exactly, it will require $\Omega(\log^2 m)$ rounds of interaction between the prover and verifier. Cormode, Mitzenmacher, and Thaler \cite{Cormode:2012:practical_streaming} gave an alternative interactive protocol for $F_0$ based on linearization, whereby the prover is more efficient in terms of running time. Their protocol requires $\log^2 m$ rounds where the verifier's space is $O(\log^2 m)$ bits and the total communication is $O(\log^3 m)$.
\par As far as we know, the only interactive protocol which uses $\log m$ rounds to compute $F_0$ is given in \cite{Cormode:2011:verifying_computation}. We note that the results stated in \cite{Cormode:2011:verifying_computation} assumed that $m=n$. We have worked out the complexity of their protocol in terms of $m$ and $n$ in Appendix~\ref{Appendix}. Restating the complexity of the $F_0$ protocol in \cite{Cormode:2011:verifying_computation} in terms of $m$ and $n$, the space of the verifier is $O\left(\log m \log n+\log^2 m\right)$ and the total communication is $O\left(\sqrt{n}\log m\left(\log n+ \log m \right)\right)$. Compared to the other protocols (e.g. $F_2$ and $\Ix$) given in \cite{Cormode:2011:verifying_computation}, the total communication is not polylogarithmic in $m$ and $n$. We briefly explain why the communication blows up to $\tilde{O}(\sqrt{n})$ in Section~\ref{Subsection:Comparison With Previous Results}.
\par Chakrabarti et al. \cite{Chakrabarti:annotations,Cormode:2010:streaming_graph_computations} studied the situation in which the prover provides a (lengthy) annotation/proof to the verifier after the data stream has ended. The verifier processes the annotation in a streaming fashion. This corresponds to randomized checking of noninteractive proofs (in the theory of interactive proofs such systems are called Merlin Arthur games, because a powerful party that is not trusted (Merlin) sends a single message to a skeptical and computationally limited verifier (Arthur)). For the exact computation of $F_0$ in this model, the help cost, $h$ and the verifier's space, $v$ are both $O(m^{2/3} \log m)$.
\par In other related work, Gur and Raz \cite{arthurmerlinstreaming} gave a Arthur-Merlin-Arthur(AMA) streaming protocol for computing $F_0$ exactly with both $h$ and $v$ being $\widetilde{O}(\sqrt{m})$ (where $\widetilde{O}$ hides a $\polylog(m,n)$ factor). Klauck and Prakash \cite{ved_hartmut_itcs13} studied a restricted interactive model where the communication between the prover and verifier has to end once the stream is already seen. Very recently, Chakrabarti et al. \cite{Chakrabarti_OIP_model} presented constant-rounds streaming interactive protocols with logarithmic complexity for several query problems, including the well studied INDEX problem.
\subsection{Previous Results and Our Techniques} \label{Subsection:Comparison With Previous Results}
\par First, we briefly describe why the protocol of \cite{Cormode:2011:verifying_computation} for computing $F_0$ fails to have total communication polylogarithmic in $m$ and $n$. It is easy to see that $F_0=\sum_{i=1}^m h\left(f_i\right)$ where $f_i:=\left|\{j\,|\,a_j=i\}\right|$ and $h:\NN\rightarrow \{0,1\}$ is given by $h(0)=0$ and $h(x)=1$ for $1\leq x \leq n$. Since $h$ depends on $n+1$ points, the degree the polynomial $\tilde{h}$, obtained via interpolation, is at most $n$, where $\tilde{h}$ agree with $h$ on $\{0,1,\cdots,n\}$. If one was to naively apply the famous sum-check protocol of Lund et al. \cite{Lund_algebraic_methods_IP}, the degree of the polynomial communicated at each round would be $O(n)$. This is even worse than the trivial protocol in which the prover either sends the frequency vector $f:=\left(f_1,\cdots,f_m\right)$ or the sorted stream, with a cost of $\tilde{O}\left(\min(m,n)\right)$. Since the proof is just a sorted stream, its correctness can be checked by standard fingerprinting techniques as described in \cite{ved_hartmut_itcs13}. One obvious way to rectify this would be to reduce the degree of the polynomial to be communicated at each round. One way to reduce the degree of $\tilde{h}$ is to remove all heavy hitters\footnote{Those items whose frequencies exceed a fraction of $n$.} from the stream, so that the degree of $\tilde{h}$ can be made small (because $h(x)$ may take any value for large $x$), which in turn means that the communication will be low. The heavy hitter protocol in \cite{Cormode:2011:verifying_computation} however uses a lot of communication just to identify all the heavy hitters, which causes the communication cost in their protocol to be high. In this work, we also reduce the degree of the polynomial to be communicated at each round. But instead of removing the heavy hitters, we write $F_0$ as a different formula. Such an approach was first used by Gur and Raz \cite{arthurmerlinstreaming} to obtain an AMA-protocol for exact $F_0$. Here, the main technical point is to replace the $OR$ function on $n$ variables which has high degree with a approximating polynomial over a smaller finite field $\FF_q$, so that this new polynomial has low degree. Such approximating polynomials were first constructed in \cite{smolensky:algebraic,Razborov86lowerbounds} to prove circuit lower bounds. The degree of the approximating polynomial $p:\FF_q^n\rightarrow \FF_q$ depends on $q$. But choosing $q$ to be small forces us to work inside the field $\FF_q$, and the arithmetic will be correct modulo $q$. Hence,  $F_0$ will be calculated modulo $q$. Note that we cannot choose $q> m$ as the approximating polynomial degree will be larger than $m$. By choosing the first $\log m$ primes, we can compute $F_0$ modulo these $\log m$ many primes with the help and verifier's cost being polylogarithmic in $m$ and $n$ (see Lemma~\ref{Lemma: F0 mod q}). This does not increase the number of rounds because all these executions can be done in parallel. The exact value of $F_0$ can be constructed by the Chinese Remainder Theorem. As a result of decreasing the degree of the polynomial, our protocol no longer has perfect completeness. By parallel repetition, the probability that a honest prover succeeds can be made close to $1$.
\par We now compare our results with previously known non-interactive and interactive protocols that compute $F_0$ exactly. For comparison purposes, we assume that $m=\theta(n)$. The results are collected in Table \ref{table:comparision of results}. We note that if we fix the number of rounds to $\log m$, our work improves the total communication from $O\left(\sqrt{m} \log^2 m \right)$ to $O\left(\log^4 m \left(\log \log m\right)^2\right)$, while only increasing the the verifier's space by a multiplicative factor of $\log \log m$. For practical purposes, the authors in \cite{Cormode:2011:verifying_computation} argue that the number of rounds in the general purpose construction of \cite{Goldwasser:2008:muggles_paper}, which is $\Omega(\log^2 m)$, may be large enough to be offputting. All the other protocols Cormode et al. \cite{Cormode:2011:verifying_computation} devise only require $\log m$ rounds. In an article in Forbes \cite{Forbes_article} in 2013, it was reported that the National Security Agency's data center in Utah will reportedly be capable of storing a yottabyte\footnote{One yottabyte is $10^{24}$ bytes.} of data. For a yottabyte-sized input, this corresponds to about $80$ rounds of interaction if one uses a protocol with $\log m$ rounds. For a protocol with $\log^2 m$ rounds, more than $6000$ rounds of interaction are needed.
\par Recently, Chakrabarti et al. \cite{Chakrabarti_OIP_model} have designed a streaming interactive protocol for the $\Ix$ function\footnote{The input stream consists of $n$ bits $x_1,\cdots,x_n$, followed by a integer $j\in [n]$. } with two messages\footnote{The first message is from the verifier to the prover and this message depends on the stream and the verifier's private randomness. The second message is from the prover to the verifier, which depends on the stream and the message received from the verifier. In general, for a $k$ message protocol, Merlin always sends the last message in the interaction.} where both space and communication are only $O(\log n \log\log n)$. Previous work gave a $\widetilde{O}(\sqrt{n})$ protocol in the this online MA model \cite{Chakrabarti:annotations}, whereas in \cite{Cormode:2011:verifying_computation}, a $\log n$ round interactive protocol with $O(\log n \log\log n)$ space and communication is given. Since for the INDEX function, there is a two message protocol with only $O(\log n \log\log n)$ complexity\footnote{The complexity of a protocol is defined as the sum of the space used and the total communication needed.}, one may ask whether a similar kind of protocol is possible for $F_0$ or other frequency moments. It is however easy to see that for $k\neq 1$, the k-th frequency moment, $F_k$ is as hard as the Disjointness function. In any online communication protocol for the Disjointness function with $2$ and $3$ messages, there is a lower bound of $\Omega(n^{1/2})$ and $\Omega(n^{1/3})$ respectively \cite{Chakrabarti_OIP_model}. Hence, it is not possible to compute $F_0$ exactly using only $1,2$ or $3$ messages with communication and space polylogarithmic in $m$ and $n$. How about using a constant number of $r$ messages, where $r \geq 4$, to get communication and space polylogarithmic in $m$ and $n$? It is believed that this is not possible: due to the recent results in \cite{Chakrabarti_OIP_model}, if Disjointness on n bits can be solved with a constant number of rounds and polylogarithmic complexity in the online one-way communication model, then the (ordinary) AM communication complexity of Disjointness will also be $polylog(n)$, which is unlikely, since Disjointness is the generic co-NP complete problem in communication complexity \cite{bfs:comm_classes}. Hence, constant round protocols ($r\geq 4$) for $F_k(k\neq1)$ with polylogarithmic complexity probably do not exist, but the current techniques in communication complexity (i.e., providing strong lower bound on the AM communication complexity of Disjointness) are not sufficient to prove this.
\begin{table}
\centering 
\begin{tabular}{c c c c } 
\hline\hline
Paper & Space & Total Communication & Number of Rounds  \\ [0.5ex] 
\hline 
\cite{Chakrabarti:annotations} & $m^{2/3} \log m$ & $m^{2/3} \log m$ & 1  \\
\cite{Cormode:2012:practical_streaming} & $\log^2 m$ & $\log^3 m$ & $\log^2 m$   \\
\cite{Cormode:2011:verifying_computation} & $\log^2 m$ & $\sqrt{m} \log^2 m$ & $\log m$   \\
This work\; \; & $\log^2 m \log \log m$ & $\log^4 m \left(\log \log m\right)^2$ & $\log m$  \\ [1ex] 
\hline 
\end{tabular}
\caption{Comparison of our protocol to previous protocols for computing the exact number of distinct elements in a data stream. The results are stated for the case where $m=\theta(n)$. The complexities of the space and the total communication is correct up to a constant.}
\label{table:comparision of results} 
\end{table}

\section{Preliminaries} \label{Section: Preliminaries}
\subsection{Data Streaming Model}
In this subsection we define our model of streaming computations with a helper/prover.

We assume that in general the input is given as a data stream $\sigma=\left<a_1,\ldots, a_n\right>$ of elements from a universe $\{1,\ldots, m\}$. The $a_i$ are sometimes referred to as symbols.

In our model we consider two parties, the prover, and the verifier. Both parties are able to access the data stream one element at a time, consecutively, and synchronously, i.e., no party can look into the future with respect to the other one. The verifier is a Turing machine that has space bounded by $\polylog(m,n)$, and processes each symbol in time $\polylog(m,n)$.

The prover is a Turing machine that has unlimited workspace, and processes each symbol in some time $T(m,n)$ that will vary from problem to problem. Ideally we want $T(m,n)$ to be $\polylog(m,n)$ as well, but this would imply immediately that the problem at hand can be solved in quasilinear time which could be too restrictive for some problems like computing the rank of a matrix.

After the stream has ended, the verifier and prover engage in a conversation to verify the correctness of some function $f(\sigma)$. The message that the prover sends to the verifier is viewed as a stream and the verifier need not store this message. He can do some computations with the message on the fly. In complexity theory, this is also known as the interactive proof model, see for instance Chapter 8 in \cite{arora&barak:cc}. In the interactive proof model in complexity theory, the prover is given unlimited power. But in our case, we want the honest prover to be able to execute our protocols efficiently. We require the verifier to run in total time $\min\{m,n\}\cdot \polylog(m,n)$, and the prover to run in time $\min\{m,n\}\cdot T(m,n)$. This makes the protocol efficient for practical delegation purposes.

Now, we are ready to define a valid protocol that verifies the correctness of some function $f(\sigma)$. We follow closely from previous works in \cite{Cormode:2011:verifying_computation,Cormode:2012:practical_streaming}.
\begin{definition}
Before seeing the stream $\sigma$, both the prover $\PP$ and verifier $\VV$ agree on a protocol to solve $f(\sigma)$. This protocol should fix all the variables that are to be used (e.g. Type of codes, size of finite fields etc.), but should not use randomness to fix these variables. If they need to choose a object from a given family according to some distribution, the verifier will choose this object and communicate the result to the prover.

After the stream ends, both $\PP$ and $\VV$ exchange some messages between each other. The message from $\PP$ to $\VV$ need not be stored but can be treated and processed as a stream. We denote the output of $\VV$ on input $\sigma$, given prover $\PP$ and $\VV$'s private randomness $\random$, by $out(\VV,\PP,\random,\sigma)$. During any phase of the interaction, $\VV$ can output $\bot$ if $\VV$ is not convinced that $\PP$'s claim is valid.

\par We say $\PP$ is a valid prover if for all streams $\sigma$,
\begin{equation*}
    Pr_R\,\left[out(\VV,\PP,\random,\sigma)=f(\sigma)\right]\geq 1-\epsilon_c.
\end{equation*}
\par We say $\VV$ is a valid verifier for $f$ if there is at least one valid prover $\PP$, and for all provers $\PP^{\prime}$ and all streams $\sigma$,
\begin{equation*}
    Pr_R\,\left[out(\VV,\PP,\random,\sigma)\notin \{f(\sigma),\bot\}\right]\leq \epsilon_s.
\end{equation*}
\end{definition}
$\epsilon_c$ is known as the completeness error, the probability that the honest prover will fail even if he follows the protocol. If $\epsilon_c=0$, we say the protocol has perfect completeness. $\epsilon_s$ is called the soundness error, that is no prover strategy will cause the verifier to output a value outside of $\{f(\sigma),\bot\}$ with probability larger than $\epsilon_s$. In this work, we take $\epsilon_c=\epsilon_s=\frac{1}{3}$. By standard boosting techniques, these probabilities can be made arbitrary close to $1$ \cite{arora&barak:cc}.
\par The main complexity measure of the protocol is the space requirement of the verifier and the total communication between the verifier and the prover. We make the following definition which takes into account these complexities.
\begin{definition} \label{def:(h,v)}
We say there is a $(h,v)$ streaming interactive protocol (SIP) with $r$ rounds that computes $f$, if there is a valid verifier $\VV$ for $f$ such that:
\begin{enumerate}
  \item $\VV$ has only access to $O(v)$ bits of working memory.
  \item There is a valid prover $\PP$ for $\VV$ such that $\PP$ and $\VV$ exchange at most $2r$ messages in total, and the sum of the length of all messages is $O(h)$ bits.
\end{enumerate}
Given any SIP, we define its complexity to be $h+v$.
\end{definition}
\par The online Merlin-Arthur(OMA) model in one where the protocol is non-interactive, in which a single message is sent from the prover to the verifier after the stream ends. As before, $\VV$ is given private randomness. Following definition~\ref{def:(h,v)}, we define
\begin{equation*}
    OMA(f)=\min\left\{\,h+v\;|\;\text{there is a $(h,v)$ online MA protocol that computes $f$}\right\}.
\end{equation*}
\subsection{Coding Theory}
We begin with some brief background from coding theory which we use in this paper. For more details of standard definitions, the reader is referred to \cite{Lint_intro_codingtheory}. A $q$-ary linear code $\code$ of length $n$ is a linear subspace of $\FF_q^n$, where $q$ is some prime power. If $\code$ has dimension $k$, then we call it a $[n,k]_q$ code. The (Hamming) distance between two codewords $x,y\in \code$, denoted by $d(x,y)$ is the number of indices $i\in[n]$ such that $x_i\neq y_i$. The distance of the code is defined as the minimum distance over all pairs of distinct codewords in $\code$. If the minimum distance of the code is $d$, we denote it by $[n,k,d]_q$. The generator matrix $G\in \FF_q^{n \times k}$ of the code is a $n$ by $k$ matrix where the column span of G gives $\code$. The relative distance of the code is $d/n$ and the rate of the code is $k/n$. A linear code is called a good code if both its relative distance and rate is at most some constant. The Reed-Solomon code is an example of a good code with alphabet size $q=n+1$. But in our case, we need the alphabet size to be much smaller than $n$. Justesen codes \cite{justesen:goodcodes} is a class of good codes with a constant alphabet size. We say that a linear code is locally logspace constructible if the $(i,j)$-entry of the generator matrix $G$ can be constructed using space $O(\log n)$. It is known that Justesen codes are locally logspace constructible~(see Lemma 3.3 of \cite{space_vs_query}).

\section{Our Result} \label{Section: Results}
\par Given a multiset presented as a stream $\sigma=\left<a_1,\cdots,a_n\right>$, where each $a_i\in [m]$, we give a interactive protocol with $\log m$ rounds which computes $F_0$ exactly. For each $j \in [m]$ and $i \in [n]$, we denote $\chi_i(j)$ to be the element indicator of element $j$ at position $i$ of the stream, i.e. $\chi_i:[m]\rightarrow \{0,1\}$ such that $\chi_i(j)=1 \Leftrightarrow a_i=j$. We can also interpret each $\chi_i:\{0,1\}^{\log m}\rightarrow \{0,1\}$ by associating each $j \in [m]$ with its binary expansion.
It is easy to see that
\begin{align*}
    F_0 &=\sum_{j=1}^m \left(\bigvee_{i=1}^n \chi_i(j)\right) \\
        &=\sum_{x_1\in\{0,1\}}\cdots\sum_{x_d\in\{0,1\}} \;\OR\left(\chi(x_1,\cdots,x_d)\right)
\end{align*}
where $d=\log m$, $\chi:\{0,1\}^d\rightarrow \{0,1\}^n$ is
\begin{equation*}
    \chi(x_1,\cdots,x_d):=\left(\chi_1(x_1,\cdots,x_d),\cdots,\chi_n(x_1,\cdots,x_d)\right).
\end{equation*}
and $\OR:\{0,1\}^n\rightarrow \{0,1\}$ is the OR function on $n$ variables.
\par Following the ideas of \cite{Lund_algebraic_methods_IP}, we consider the low degree extension of $\chi_i$ over a larger field. Let $q$ be a prime and $\lambda$ an integer to be determined later. We extend the domain of $\chi_i$ from $\FF_2^d$ to $\FF_{q^\lambda}^d$. If we denote $\theta_{\sigma}:[m]\rightarrow \FF_2^d$ where $\theta_{\sigma}(i)=\left(a_i^{(1)},\cdots,a_i^{(d)}\right)$ is the binary expansion of $a_i$,
then the extension $\widetilde{\chi_i}:\FF_{q^\lambda}^d\rightarrow \FF_{q^\lambda}$ is given by
\begin{equation}  \label{Eq: chi as a polynomial}
    \widetilde{\chi_i}(x_1,\cdots,x_d):=\prod_{j=1}^d \left[\left(2a_i^{(j)}-1\right)x_j + \left(1-a_i^{(j)}\right)\right].
\end{equation}
Note that $\widetilde{\chi_i}(x_1,\cdots,x_d)=\chi_i(x_1,\cdots,x_d)$ for all $x\in \FF_2^d$. Similarly, define
$\widetilde{\chi}:\FF_{q^\lambda}^d\rightarrow \FF_{q^\lambda}^n$ in the natural way:
\begin{equation*}
    \widetilde{\chi}(x_1,\cdots,x_d):=\left(\widetilde{\chi_1}(x_1,\cdots,x_d),\cdots,\widetilde{\chi_n}(x_1,\cdots,x_d)\right).
\end{equation*}
With this notation,
\begin{equation} \label{Eq: FO w/o randomization}
    F_0=\sum_{x_1\in\{0,1\}}\cdots\sum_{x_d\in\{0,1\}} \;\OR\left(\widetilde{\chi}(x_1,\cdots,x_d)\right).
\end{equation}
Running the sum-check protocol naively to~(\ref{Eq: FO w/o randomization}) would require the prover to send a degree $n$ polynomial at each round. We replace the OR function in~(\ref{Eq: FO w/o randomization}) with a low degree polynomial which approximates the OR function with high probability. This idea was first introduced in \cite{smolensky:algebraic,razborov_lowerbound_circuit} and was also used in \cite{arthurmerlinstreaming} to obtain an AMA protocol for exact $F_0$.
\begin{lemma} \label{lemma:randomized approxiamtion of OR}
Using $O(L \log n)$ bits of randomness, we can construct a polynomial $p:\FF_q^n\rightarrow \FF_q$ of individual degree at most $L(q-1)$, such that for every $x\in \{0,1\}^d$,
\begin{equation*}
    Pr\left[p\left(\widetilde{\chi}(x_1,\cdots,x_d)\right)=\OR\left(\widetilde{\chi}(x_1,\cdots,x_d)\right)\right]
         \geq 1-\frac{1}{6m\log m},
\end{equation*}
where $L$ is the least integer such that
\begin{equation} \label{Eq: choice of L}
    \left(\frac{2}{3}\right)^L\leq\frac{1}{6m\log m}.
\end{equation}
\end{lemma}
\begin{proof}
Start with a $[\zeta n, n, \frac{1}{3}\zeta n]_q$-linear code $\code$, where $\zeta>1$ is a constant to be chosen such that $\code$ exist\footnote{Justesen codes \cite{justesen:goodcodes} are one example of a family of codes which have both constant relative distance and constant rate.}. Let $G$ be the generator matrix of $\code$. Choose uniformly at random $\alpha_1,\cdots,\alpha_L \in [\zeta n]$ where $L$ is the least integer that satisfies~(\ref{Eq: choice of L}) and define
\begin{equation*}
    p(x_1, \cdots, x_n):=1-\prod_{i=1}^L \left[1-\left((Gx)_{\alpha_i}\right)^{q-1}\right].
\end{equation*}
It is easy to see that the individual degree of $p$ is at most $L(q-1)$. By properties of the code $\code$, for any $x \in \{0,1\}^n$,
\begin{equation*}
    \Pr_{\alpha_1,\cdots,\alpha_L}\left[p(x)\neq \bigvee _i x_i\right] \leq \left(\frac{2}{3}\right)^L \leq \frac{1}{6m\log m}.
\end{equation*}
\end{proof}
We note that $L=O(\log m)$.
\par Since $\FF_{q^\lambda}$ can be viewed as a vector space over $\FF_q$, we can view $p:\FF_q^n\rightarrow \FF_q$ as $\widetilde{p}:\FF_{q^\lambda}^n\rightarrow \FF_{q^\lambda}$, by applying $p$ componentwise. By the union bound, the probability that
\begin{equation} \label{Eq: prob that p represents OR function}
     Pr\left[F_0  \pmod q =\sum_{x_1\in\{0,1\}}\cdots\sum_{x_d\in\{0,1\}} \;\widetilde{p}\left(\widetilde{\chi}(x_1,\cdots,x_d)\right)\right] \geq 1-\frac{1}{6\log m}.
\end{equation}
\par We first give a interactive protocol to compute $F_0  \pmod q$ with high probability. Let $q\leq 2\log m \log\log m +2$ be a prime and $\lambda$ be the smallest integer such that $q^{\lambda-1} \geq 6Ld\log m$. Before observing the stream, the prover and verifier agree on the code $\code$ as in Lemma~\ref{lemma:randomized approxiamtion of OR}. The verifier chooses $O(L \log n) $ random bits to define the polynomial $\widetilde{p}$ and sends this randomness to the prover. The verifier chooses randomly $r \in \FF_{q^\lambda}^d$ and computes $\widetilde{p}\left(\widetilde{\chi}(r_1,\cdots,r_d)\right)$ in a streaming fashion. We now illustrate how the verifier computes $\widetilde{p}\left(\widetilde{\chi}(r_1,\cdots,r_d)\right)$ given a one-pass over the stream without storing the whole input.
\par For $1\leq i \leq n$, let $y_i=\left(y_i^{(1)},\cdots,y_i^{(\lambda)}\right)$ where $y_i=\widetilde{\chi_i}(r_1,\cdots,r_d)$ and each $y_i^{(j)}\in \FF_q$. Each $y_i$ can be computed when $a_i$ is seen in the stream using (\ref{Eq: chi as a polynomial}).

Then
\begin{align*}
   \widetilde{p}\left(\widetilde{\chi}(r_1,\cdots,r_d)\right) &=\widetilde{p}\left(y_1,\cdots, y_n\right) \\
     &=\left(p\left[\begin{array}{c}
                y_1^{(1)} \\
                \vdots \\
                y_n^{(1)}
              \end{array}\right],\quad \cdots, \quad p\left[\begin{array}{c}
                y_1^{(\lambda)} \\
                \vdots \\
                y_n^{(\lambda)}
              \end{array}\right]
     \right)
\end{align*}
We show how to compute $p\left[\begin{array}{c}
                y_1^{(j)} \\
                \vdots \\
                y_n^{(j)}\end{array}\right]$ for any $1 \leq j \leq \lambda$ in a streaming fashion. For $1 \leq j \leq \lambda$ and $1\leq i \leq L$, $\VV$ needs to compute
\begin{align} \label{Eq: B matrix}
   B_{ij}&=\left(G\left[\begin{array}{c}
                y_1^{(j)} \\
                \vdots \\
                y_n^{(j)}
              \end{array}\right]\right)_{\alpha_i} \\
         &=\left(G\left[\begin{array}{c}
                y_1^{(j)} \\
                0 \\
                \vdots \\
                0
              \end{array}\right]\right)_{\alpha_i}+\cdots+\left(G\left[\begin{array}{c}
                0 \\
                \vdots \\
                0 \\
                y_n^{(j)}
              \end{array}\right]\right)_{\alpha_i}  \nonumber
\end{align}
This can be done given a one pass over the stream $\sigma$, each time $\VV$ observes a new entry $a_k$, he updates
\begin{equation*}
    B_{ij}\leftarrow B_{ij} + y_k^{(j)} \cdot g_{\alpha_i,k}.
\end{equation*}
Note that the computation of $y_k^{(j)}$ depends only on $a_k$ and the verifier need not store matrix $G$. Upon observing entry $a_k$, only the $\alpha_1,\cdots,\alpha_L$ entries of the k-th column of $G$ are relevant. Since $G$ is locally logspace constructible, each $g_{\alpha_i,k}$ can be constructed in $O(\log n)$ space. After the stream has ended, the verifier computes $p\left[\begin{array}{c}
                y_1^{(j)} \\
                \vdots \\
                y_n^{(j)}\end{array}\right]$
using
\begin{equation*}
      p\left[\begin{array}{c}
                y_1^{(j)} \\
                \vdots \\
                y_n^{(j)}
              \end{array}\right] =1-\prod_{i=1}^L \left(1-B_{ij}^{q-1}\right).
\end{equation*}
\par After the stream ends, the verification protocol proceeds in $d$ rounds to compute $F_0  \pmod q$ with probability at least $1-\frac{1}{6\log m}$. In the first round, the prover sends a polynomial $g_1(X_1)$ which is claimed to be
\begin{equation*}
    g_1(X_1)=\sum_{x_2\in\{0,1\}}\cdots\sum_{x_d\in\{0,1\}} \;\widetilde{p}\left(\widetilde{\chi_1}\left(X_1,x_2,\cdots,x_d\right),\cdots,\widetilde{\chi_n}
    \left(X_1,x_2\cdots,x_d\right)\right).
\end{equation*}
The polynomial $g_1(X_1)$ has degree $L(q-1)$ which can be described in $O(Lq\log q^\lambda)$ bits. The verifier need not store $g_1(X_1)$ but just need to compute $g_1(r_1)$, $g_1(0)$ and $g_1(1)$, which can be done in a streaming fashion. Note that if the prover is honest, then
\begin{equation} \label{Eq: F0 mod q from IP protocol}
    F_0  \pmod q=g_1(0)+g_1(1).
\end{equation}
In round $2 \leq j \leq d-1$, the verifier sends $r_{j-1}$ to the prover who then sends the polynomial $g_j(X_j)$, which is claimed to be
\begin{equation} \label{Eq:g_j as a polynomial}
\begin{split}
     g_j(X_j)=\sum_{x_{j+1}\in\{0,1\}}\cdots\sum_{x_d\in\{0,1\}} \; \widetilde{p}(&\widetilde{\chi_1}\left(r_1,\cdots,r_{j-1},X_j,x_{j+1},\cdots,x_d\right),\cdots \\
      \cdots, &\widetilde{\chi_n}\left(r_1,\cdots,r_{j-1},X_j,x_{j+1},\cdots,x_d\right))
\end{split}
\end{equation}
The verifier computes $g_j(r_j)$, $g_j(0)$ and $g_j(1)$ and proceeds to the next round only if the degree of $g_j$ is at most $L(q-1)$ and
\begin{equation*}
    g_{j-1}(r_{j-1})=g_j(0)+g_j(1).
\end{equation*}
In the final round, the verifier sends $r_{d-1}$ to the prover who then sends the polynomial $g_d(X_d)$, which is claimed to be
\begin{equation*}
      g_d(X_d)=\widetilde{p}\left(\widetilde{\chi_1}\left(r_1,\cdots,r_{d-1},X_d\right),\cdots, \widetilde{\chi_n}\left(r_1,\cdots,r_{d-1},X_d\right)\right).
\end{equation*}
The verifier only accepts that~(\ref{Eq: F0 mod q from IP protocol}) is computed correctly if $g_d$ is of the correct degree, $g_{d-1}(r_{d-1})=g_d(0)+g_d(1)$ and $g_d(r_d)=\widetilde{p}\left(\widetilde{\chi}(r_1,\cdots,r_d)\right)$. \\
Next, we show that if the prover is dishonest, the verifier will reject the claimed value of $F_0  \pmod q$ with high probability.
\begin{lemma} \label{Lemma: correctness of sum check protocol}
In the case of the honest prover, the verifier will accept the wrong value of $F_0  \pmod q$ with probability at most $\frac{1}{6\log m}$. If
\begin{equation} \label{Eq:d-sum of approx OR function}
    \sum_{x_1\in\{0,1\}}\cdots\sum_{x_d\in\{0,1\}} \;\widetilde{p}\left(\widetilde{\chi}(x_1,\cdots,x_d)\right)
\end{equation}
correctly represents $F_0  \pmod q$ and if the prover cheats by sending some polynomial which does not need the requirements of the protocol, the verifier will accept with probability at most $\frac{L(q-1)d}{q^\lambda}$.
\end{lemma}
\begin{proof}
In the case of a honest prover, since the interactive protocol always evaluates~(\ref{Eq:d-sum of approx OR function}) correctly, the prover will fail in the case that the approximating polynomial $\widetilde{p}$ does not represent the OR function. By~(\ref{Eq: prob that p represents OR function}), the probability that the honest prover will fail is at most $\frac{1}{6 \log m}$. \\
\par For the case of the dishonest prover, the argument proceeds inductively from the $d$-th round to the first round. Indeed, if $g_d$ is not as claimed, by the Schwartz-Zippel lemma,
\begin{equation*}
    Pr\left[g_d(r_d)=\widetilde{p}\left(\widetilde{\chi}(r_1,\cdots,r_d)\right)\right]\leq \frac{L(q-1)}{q^\lambda}.
\end{equation*}
By induction, suppose for $1\leq j \leq d-1$ that the verifier is convinced that $g_{j+1}(X_{j+1})$ is correct with high probability. He can verify the correctness of $g_j(X_j)$ with high probability; since
\begin{equation*}
    \begin{split}
     g_{j+1}(X_{j+1})=\sum_{x_{j+2}\in\{0,1\}}\cdots\sum_{x_d\in\{0,1\}} \; \widetilde{p}(&\widetilde{\chi_1}\left(r_1,\cdots,r_j,X_{j+1},x_{j+2},\cdots,x_d\right),\cdots \\
      \cdots, &\widetilde{\chi_n}\left(r_1,\cdots,r_j,X_{j+1},x_{j+2},\cdots,x_d\right)),
\end{split}
\end{equation*}
it is easy to see by~(\ref{Eq:g_j as a polynomial}) that $g_j(r_j)=g_{j+1}(0)+g_{j+1}(1)$. By the Schwartz-Zippel lemma,
\begin{equation*}
\begin{split}
    Pr\;\Big[\hat{g_j}(r_j)=g_{j+1}(0)+g_{j+1}(1)\quad \text{where}\; \\
    \hat{g_j}(X_j)\neq
     \sum_{x_{j+1}\in\{0,1\}}\cdots\sum_{x_d\in\{0,1\}} \; \widetilde{p}(&\widetilde{\chi_1}\left(r_1,\cdots,r_{j-1},X_j,x_{j+1},\cdots,x_d\right),\cdots \\
      \cdots, &\widetilde{\chi_n}\left(r_1,\cdots,r_{j-1},X_j,x_{j+1},\cdots,x_d\right))
 \Big] \leq \frac{L(q-1)}{q^\lambda}.
 \end{split}
\end{equation*}
For the cheating prover to succeed, he has to give
\begin{equation*}
    \hat{g_1}(X_1)\neq \sum_{x_2\in\{0,1\}}\cdots\sum_{x_d\in\{0,1\}} \;\widetilde{p}\left(\widetilde{\chi_1}\left(X_1,x_2,\cdots,x_d\right),\cdots,\widetilde{\chi_n}
    \left(X_1,x_2\cdots,x_d\right)\right)
\end{equation*}
and either that in some round $j+1$ (for some $1\leq j \leq d-1$), when the verifier reveals $r_j$, it should satisfy
\begin{equation*}
    \hat{g_j}(r_j)=g_{j+1}(0)+g_{j+1}(1)
\end{equation*}
or $\hat{g_d}(r_d)=\widetilde{p}\left(\widetilde{\chi}(r_1,\cdots,r_d)\right)$ in the final round. By the union bound,
\begin{equation*}
    Pr\left[ \hat{g_1}(X_1)\neq g_1(X_1)\; \text{and the verifier accepts}\right]\leq \frac{L(q-1)d}{q^\lambda}\leq \frac{1}{6\log m}.
\end{equation*}
\end{proof}
\noindent\textbf{Analysis of space and communication.} We now analyse the space needed by the verifier and the total communication between the prover and verifier over the $\log m$ rounds to verify $F_0  \pmod q$. First, let us look at the space complexity of the verifier. He needs to store $\alpha_1,\cdots,\alpha_L$ which will take $O(\log m \log n)$ bits of space. With $O(\log m \log\log m)$ bits of space, the verifier can compute $\widetilde{p}\left(\widetilde{\chi}(r_1,\cdots,r_d)\right)$ when observing the stream. Note during the interaction with the prover after the stream ends, at each round $1\leq j \leq d$, the verifier need not store the polynomial $g_j(X_j)$ but only need to evaluate $g_j$ at a constant number of points. Hence, the space complexity of the verifier is $O\left(\,\log m  \left[\,\log n + \log\log m\,\right]\right)$ bits.
\par We now bound the total communication between the prover and verifier. The verifier needs to communicate $\alpha_1,\cdots,\alpha_L$ and $r_1,\cdots,r_{d-1}$ to the prover, with cost $O(\log m \log n)$ and $O(\log m \log \log m)$ respectively. The prover, who needs to send $g_1(X_1),\cdots,g_d(X_d)$, uses $O(dLq \log q^{\lambda})=O(q\cdot \log^2 m \log\log m)$ bits to communicate all these polynomials. Hence, the total communication is $O\left(\,\log m\left(\log n +q \log m \log\log m\right)\right)$ bits. We summarize our result below.
\begin{lemma} \label{Lemma: F0 mod q}
There exist an $(h,v)$ SIP with $\log m$ rounds with
\begin{align*}
    h&=\log m\left(\log n +q \log m \log\log m\right), \\
    v&=\log m \left(\,\log n + \log\log m\,\right)
\end{align*}
that computes $F_0  \pmod q$, where the completeness error
\begin{equation*}
    \epsilon_c=\frac{1}{6 \log m}
\end{equation*}
and the soundness error
\begin{equation*}
    \epsilon_s=\frac{1}{3 \log m}.
\end{equation*}
for any prime $q\leq 2\log m \log\log m +2$.
\end{lemma}
\noindent\textbf{Computing $F_0$ exactly.} Lemma~\ref{Lemma: F0 mod q} gives us an streaming interactive protocol to verify the correctness of $F_0  \pmod q$ with high probability for any prime $q\leq 2\log m \log\log m +2$. Now, we show how the prover can verify $F_0$ with high probability. Let $Q=\{q_1,\cdots,q_{\log m}\}$ be the first $\log m$ primes. Note that $q_{\log m}\leq 2\log m \log\log m +2$ for all $m\geq 2$ \cite{bach_book_algo_number_theory_vol1} and
\begin{equation*}
    \prod_{i=1}^{\log m} q_i >m.
\end{equation*}
So, the verifier will compute $F_0  \pmod {q_i}$ for $i=1,\cdots,\log m$. Note that this can be done in parallel and will cause the working space of the verifier and the total communication to increase, but the number of rounds is still $\log m$. By using the Chinese remainder theorem, the verifier can compute $F_0$ exactly given $F_0  \pmod {q_i}$ for $i=1,\cdots,\log m$. By the union bound, the completeness and soundness error are $1/6$ and $1/3$ respectively.

In the preprocessing phase (even before seeing the data), the verifier and prover agree on a constant $\zeta>0 $ such that the linear code $\code_i:=[\zeta n, n, \frac{1}{3}\zeta n]_{q_i}$ exists for all $1 \leq i \leq \log m $. Note that the same $\alpha_1,\cdots,\alpha_L$ can be used to define the polynomial $\widetilde{p_i}:\FF_{q_i^\lambda}^n\rightarrow \FF_{q_i^\lambda}$ for each $1 \leq i \leq \log m $. For each $1 \leq i \leq \log m $, the verifier needs to choose uniformly at random  $r^{(i)} \in \FF_{q_i^\lambda}^d$ and compute $\widetilde{p}\left(\widetilde{\chi}\left(r^{(i)}\right)\right)$. This can be done in space $O\left(\log^2 m \log\log m\right)$. Hence, the total space used by the verifier is $O(\log m \left(\,\log n + \log m \log\log m\,\right))$.

To bound the total communication, we need the following fact: Let $p_n$ be the $n^{th}$ prime, then it is known \cite{bach_book_algo_number_theory_vol1} that $\sum_{i=1}^{n} p_i=\Theta(n^2 \log n)$ for all $n\geq 2$. Hence, the total communication is
\begin{equation*}
   O\left( \log m \log n + (\log^2 m \log\log m)\sum_{i=1}^{\log m} q_i \right)=O\left(\log m \log n+\log^4 m (\log\log m)^2 \right).
\end{equation*}
\noindent\textbf{Running time of the verifier.} First, we analyse the processing time of each symbol seen in the stream. We suppose it takes unit time to add and multiply two field elements from $\FF_{q^\lambda}$. For each symbol $a_k$ seen, the verifier needs to compute $\widetilde{\chi_k}\left(r^{(q)}\right)$ where $r^{(q)} \in \FF_{q^\lambda}^d$ for each $q \in Q$. From~(\ref{Eq: chi as a polynomial}), it is easy to see that the verifier needs $O(d)=O(\log m)$ time to compute $\widetilde{\chi_k}\left(r^{(q)}\right)$ for each $q\in Q$. Hence the total time taken by the verifier to process each symbol is $O(\log^2 m)$. After this, the verifier can discard $a_k$ and only need to update matrices $B$ for each $q\in Q$ (See~(\ref{Eq: B matrix}) for the definition of matrix $B$). In another workstation, the verifier can update matrices $B$ using~(\ref{Eq: B matrix}) after computing $\widetilde{\chi_k}\left(r^{(q)}\right)$ for each $q\in Q$. Note that after $\widetilde{\chi_k}\left(r^{(q)}\right)$ is computed, the updating of $B$ does not require $a_k$ anymore.

We summarize our results below.
\begin{theorem} \label{Thm:F0 protocol complexity}
There exist an $(h,v)$ SIP with $\log m$ rounds with
\begin{align*}
    h&=\log m \left(\,\log n +\log^3 m (\log\log m)^2 \,\right), \\
    v&=\log m \left(\,\log n + \log m \log\log m\,\right)
\end{align*}
that computes $F_0$ exactly, where the completeness and soundness error are $1/6$ and $1/3$ respectively. The update time for the verifier per symbol received is $O(\log^2 m)$.
\end{theorem}

\section{Conclusions and Open Problems}

We have shown that there is a streaming interactive protocol with $\log m$ rounds to compute $F_0$ exactly using space and communication polylogarithmic in $m$ and $n$. This improves the previous work in~\cite{Cormode:2011:verifying_computation} which also gave a $\log m +1$ protocol but the total communication of their protocol was $\widetilde{O}(\sqrt{n})$. 
\par In this section, we assume that $m=\theta(n)$, to simplify the statement of bounds.
An open problem that remains is whether we can obtain an interactive protocol with $\log m$ rounds and $\polylog(m)$ space and communication for $F_{\infty}$, the number of times the most frequent item appears in the data stream. It is known that a constant approximation of $F_{\infty}$ requires $\Omega(m)$ space in the standard model where there is no prover. So even approximating $F_{\infty}$ with $\log m$ rounds and $\polylog(m)$ complexity is interesting. Again computing $F_\infty$ in in NC and polylogarithmic rounds can be achieved. The authors of \cite{Cormode:2011:verifying_computation} describe an interactive protocol for exactly computing $F_{\infty}$ with $\log m$ rounds with communication $\widetilde{O}(\sqrt{m})$ and polylogarithmic space.
\par Another interesting theoretical problem would be to obtain non-trivial lower bounds on the Arthur-Merlin(AM) communication complexity
of the Disjointness function (refer to \cite{klack_AM_games_CC} for the definition of the AM model in communication complexity). Proving any superlogarithmic lower bounds on the AM complexity of Disjointness will rule out constant round streaming interactive protocols for $F_k(k\neq 1)$ with polylogarithmic complexity.
\par For the online MA model, we conjecture that the protocol given in \cite{Chakrabarti:annotations} for exact $F_0$ with complexity $O(m^{2/3} \log m)$ is tight, up to logarithmic factors. In this restricted online MA model, it might be easier to prove lower bounds larger than $\sqrt{m}$, as compared to proving lower bounds in the general Merlin-Arthur model. Also, in the online MA model, any protocol that approximates $F_k$ up to a constant factor requires $hv=\Omega(m^{1-5/k})$ \cite{Chakrabarti:annotations}. Since it is known that for $k\geq 3$, approximating $F_k$ in the standard model requires $\Omega(m^{\alpha})$ space for some constant $\alpha>0$ \cite{yoseff:datastream}, it is an interesting open problem whether approximating $F_3, F_4$ and $F_5$ is easy or hard in the online MA model.
\bibliographystyle{plain}
\bibliography{references}
\appendix
\section{$F_0$ protocol of \cite{Cormode:2011:verifying_computation}} \label{Appendix}
\par We note that the protocol given in \cite{Cormode:2011:verifying_computation} assumes that $n=O(m)$. We present their protocol for completeness sake, stating the complexity results in terms of $m$ and $n$.
\begin{theorem} \label{Thm:F0 protocol for CMT11}
There is a $(\,\sqrt{n} \log m (\log m +\log n)\,,\,\log m(\log m +\log n)\,)$ SIP with $\log m +1$ rounds to compute $F_0$.
\end{theorem}
Following the discussion in Section~\ref{Subsection:Comparison With Previous Results}, we remove the heavy hitters from the stream to reduce the degree of the polynomial. The $\phi$-heavy hitter protocol described below lists all the items $i \in [m]$ such that $f_i > T:=\phi n$. The verifier need not store all the heavy hitters in his memory.
\begin{lemma}
There is a $(\,\phi^{-1} \log m (\log m +\log n)\,,\,\log m(\log m +\log n)\,)$ SIP with $\log m +1$ rounds that identifies all the $\phi$-heavy hitters in a data stream.
\end{lemma}
\begin{proof}
Consider a binary tree $\Tree$ of depth $\log m$ where the value of the $i$-th leaf is $f_i$. For any node $v \in V_{\Tree}$, denote $L(v)$ to be the set of leaves of the subtree rooted at $v$ and let $p(v)$ to be the parent of $v$. For every node $v \in V_{\Tree}$, we denote its value by $\widehat{f}(v):=\sum_{i \in L(v)} f_i$. We denote the witness set $W \subseteq V_{\Tree}$ which consists of all leaves $l$ with $\widehat{f}(l)>T$ and all nodes $v$ which satisfy $\widehat{f}(v)\leq T$ and $\widehat{f}(p(v)) > T$. This witness set ensures that no heavy hitters are omitted by the prover. Label the nodes of $\Tree$ in some canonical order from $\{1,2,\cdots,2m-1\}$. Let $x\in\{0,1\}^{2m-1}$ be the indicator vector for $W$, i.e. $x_j=1$ if and only if the $j$-th node of $\Tree$ belongs to $W$.
The prover gives the set $W$ together with the claimed frequency $f^{\ast}(w)$ for each $w \in W$. If $w \notin W$, then define $f^{\ast}(w)=0$. The verifier needs to check that the set $W$ does cover the whole universe and that $f^{\ast}(w)=\widehat{f}(w)$ for all $w \in W$. This is equivalent to checking that
\begin{equation} \label{Eq: heavy hitter protocol}
    \sum_{j=1}^{2m-1} x_j\left(f^{\ast}(j)-\widehat{f}(j)\right)^2=0.
\end{equation}
Running the sum-check protocol of \cite{Lund_algebraic_methods_IP} to (\ref{Eq: heavy hitter protocol}) will require $\log m +1$ rounds of interaction. We have to choose a prime $q>(2m-1)n^2$ which will require $O(\log n +\log m)$ bits to represent $q$. At each level of the binary tree, there can be at most $2\phi^{-1}$ nodes that belong to $W$, which implies that $|W|=O(\phi^{-1} \log m)$. In each round of the sum-check protocol, the prover communicates a polynomial of degree at most $3$. Hence the total communication is dominated by $O(\phi^{-1} \log^2 m +\phi^{-1} \log m \log n)$. \\
The verifier need to pick a random point $r\in \FF_q^{\log m +1}$ and needs to evaluate $\widetilde{x}(r)$, $\widetilde{f^{\ast}}(r)$ and $\widetilde{\widehat{f}}(r)$, where $\widetilde{x}$, $\widetilde{f^{\ast}}$ and $\widetilde{\widehat{f}}$ are the multilinear extensions of $x$, $f^{\ast}$ and $\widehat{f}$ respectively. As described in~\cite{Cormode:2011:verifying_computation}, $\widetilde{x}(r)$, $\widetilde{f^{\ast}}(r)$ and $\widetilde{\widehat{f}}(r)$ can be calculated in a single pass over the stream. Hence the space complexity of the verifier is $O(\log m(\log m+\log n))$.
\end{proof}
\par The protocol of $F_0$ proceeds as follows: The verifier removes all the $\phi$-heavy hitters in the stream with the help of the prover. This gives rise to a new stream $\widetilde{\sigma}$ where the frequency of each element is at most $\phi n$. The sum check protocol~\cite{Lund_algebraic_methods_IP} is then applied to
\begin{equation} \label{Eq:sum check protocol on derived stream for F0}
    F_0=\sum_{i=1}^m h\left(f'_i\right)
\end{equation}
where $f'_i$ is the frequency of item $i$ in the derived stream $\widetilde{\sigma}$ and $h:\NN\rightarrow \{0,1\}$ is given by $h(0)=0$ and $h(x)=1$ for $1\leq x \leq \phi n$. We can work over the finite field $\FF_p$, where $p\geq \max\{m,n\}$. Running the sum check protocol on~(\ref{Eq:sum check protocol on derived stream for F0}) requires $\log m$ rounds of interaction which will give a total communication of $O(\phi n(\log m + \log n)\log m)$. The space required by the verifier is $O(\log m(\log m+\log n))$. Note that the heavy hitter protocol and the sum check protocol of~(\ref{Eq:sum check protocol on derived stream for F0}) can be carried out in parallel. By choosing $\phi=\frac{1}{\sqrt{n}}$, we get Theorem~\ref{Thm:F0 protocol for CMT11}.
\end{document}